\documentclass[twocolumn, twocolappendix]{aastex63}
\usepackage{soul}
\usepackage{todonotes}
\usepackage{ulem}
\usepackage{amsmath}
\usepackage{amssymb}

\definecolor{DarkerGreen}{rgb}{0.0,0.7,0.1}

\newcommand{\indep}{{\buildrel{\rm indep}\over\sim}}

\received{June 1, 2019}
\revised{January 10, 2019}
\accepted{\today}
\submitjournal{ApJS}

\shorttitle{Six Maxims for Sound Astronomical Data Analysis}
\shortauthors{Tak et al.}

\begin{document}

%\title{Statistical Maxims for Sound Astronomical Data Analysis}
\title{Six Maxims of Statistical Acumen for Astronomical Data Analysis}
% Tuned by the Stories Behind the Data
%\dvd{Four Maxims for}

\correspondingauthor{Hyungsuk Tak (tak@psu.edu)}
%

%\author{Hyungsuk Tak}
%\author[]{David A van Dyk}

\author[0000-0003-0334-8742]{Hyungsuk Tak}
\affiliation{Departments of Statistics, Astronomy \& Astrophysics, Institute for Computational and Data Sciences, Pennsylvania State University \\
%1667 K Street NW, Suite 800 \\
University Park, Pennsylvania 16802, USA}

\author[0000-0002-9516-8134]{Yang Chen}
\affiliation{Department of Statistics, University of Michigan \\
%1667 K Street NW, Suite 800 \\
Ann Arbor, Michigan 48109, USA}

\author[0000-0002-3869-7996]{Vinay L. Kashyap}
\affiliation{Harvard Smithsonian Center for Astrophysics\\
Cambridge, Massachusetts 02138, USA} 
%\email{vkashyap@cfa.harvard.edu}

\author[0000-0001-9846-4417]{Kaisey S. Mandel}
\affiliation{Institute of Astronomy and Kavli Institute for Cosmology, University of Cambridge, Cambridge CB3 0HA, UK}
%\email{kmandel@ast.cam.ac.uk}
\affiliation{Statistical Laboratory, DPMMS, University of Cambridge, Wilberforce Road, Cambridge, CB3 0WB, UK}

\author[0000-0003-3687-0385]{Xiao-Li Meng}
\affiliation{Department of Statistics, Harvard University\\
Cambridge, Massachusetts 02138, USA}
%\email{meng@stat.harvard.edu}

\author[0000-0002-0905-7375]{Aneta Siemiginowska}
\affiliation{Harvard Smithsonian Center for Astrophysics\\
Cambridge, Massachusetts 02138, USA} 
%\email{asiemiginowska@cfa.harvard.edu}

\author[0000-0002-0816-331X]{David A. van Dyk}
\affiliation{Statistics Section, Department of Mathematics, Imperial College London \\
%1667 K Street NW, Suite 800 \\
London SW7 2AZ, UK}
%\email{dvandyk@imperial.ac.uk}

%\dvd{I have revised the abstract. Now 175 words.}
\begin{abstract}

The acquisition of complex astronomical data  is accelerating, especially with newer telescopes producing ever more large-scale surveys. The increased quantity, complexity, and variety of astronomical data demand a parallel increase in skill and sophistication in developing, deciding, and deploying statistical methods. Understanding limitations and appreciating nuances in statistical and machine learning methods and the reasoning behind them is essential for improving data-analytic proficiency and acumen. Aiming to facilitate such improvement in astronomy, we delineate cautionary tales in statistics via six maxims, with examples drawn from the astronomical literature.
Inspired by the significant quality improvement in business and manufacturing processes by the routine adoption of Six Sigma, we hope the routine reflection on these Six Maxims will improve the quality of both data analysis and scientific findings in astronomy.
%\dvd{say}
%\dvd{I don't understand this quotation. Do you mean ``garbage in, garbage out'' or ``garbage model, garbarge out''? or maybe: To steer clear of ``garbage out, because of a garbage model,'' to coin a clunky phrase,}
%(which can also be applicable to machine-learning-based data analysis?)
%\sout{We set out  typical checks and validations that statisticians do when confronted with data analysis problems, arranged in a form applicable to astronomical analysis.  While many of the maxims will be familiar in a general sense to astronomers, we describe several concrete exemplar templates drawn from the astronomical literature which can help inform the set up for future analyses. We hope that researchers can easily check these maxims as part of their daily data-analysis routine and that they improve the quality of both data analysis and scientific findings in astronomy.} 

\end{abstract}

%--

%editorials, notices --- 
%miscellaneous --- catalogs --- surveys \dvd{can we review these?}
%

%\citep{box1976}, 
%\begin{center}
%\emph{``All models are wrong.''}
%\end{center}
%Later, he appended the aphorism into its most  widely known form

%: A motivating example
\section{Introduction} \label{sec1}

Although data science aims
%is defined 
to address the myriad challenges arising from the entire life cycle of data \citep{Wing2019Data}, there are a number of unique, or at least unusual, characteristics of astronomical data that demarcate the
statistical
%\dvd{statistical} 
challenges in astronomy and affect our approach to analyzing  astronomical data.  
First, astronomical observations are not obtained from designed experiments in the traditional sense. % of the concept 
There are no experimental settings  that the astrophysical researcher compares by controlling conditions, such as  treatment versus placebo. Consequently, observations are not exactly repeatable in the sense that  observational conditions, instrumental properties, or the astrophysical phenomena themselves are changing over time. For instance, we cannot observe the same supernova explosion multiple times under different controlled conditions.
%\dvd{I don't follow the non-repeatable part. They are not repeatable because if you look at something at a later date, it may have changed. But this is true in experiments too.} 
%\vlk{In physics experiments though you can set up the conditions to be repeatable, e.g., multiple LHC runs.  But once a supernova goes off, that's it for that star.  Astronomy does not allow knobs to be turned except in crude natural experiments, like looking at G type stars of $\approx$5~Gyr age but with different compositions (say) to explore how the Sun would behave in a different circumstance.} 
Second, calibration (e.g., of instruments) is a crucial step of the observation process because it allows us to connect the observed signals to the underlying physics \citep[see, e.g.,][]{2015JATIS...1d7001G}. 
Unfortunately, calibration is never exact and thus adds uncertainty to the final analysis \citep[e.g.,][]{2011ApJ...731..126L}. 
%\vlk{if we cite something here, it would have to be a long list.  Lee et al 2011 is an obvious one.  Villanueva et al 2021 Nature Astronomy (phosphine).  Payne et al 2020 PhysRevD (GW wave systematics).  Guainazzi et al 2015 JATIS (Xray).  Partridge et al 2016 ApJ (Planck). Marshall 2021 (polarization) need more}.
Third, sparsity is inevitable even with big data, in a sense that researchers are always interested in the most distant objects and the faintest signals which are studied using small subsets of the full data. For instance, new classes of astronomical phenomena are rarely first discovered as bright sources, but are often among the most interesting scientifically. 
Fourth, observed astronomical objects are at different stages of their life cycles and evolve on different time scales, 
all of which
%\dvd{all of which} 
%that 
are much larger than we can observe. This characteristic can help us understand long time scales via a population study, but homogeneity and completeness of the data may become an issue.
% that are not accessible by direct observation
Finally,   measurement error uncertainties are heteroscedastic and are often given as constants along with the data  \citep{feigelson1998statistical, feigelson2021twenty}. 

Taken together, these unusual characteristics can lead to challenges, especially when using off-the-shelf data-analytic tools to analyze astronomical data \citep{Siemiginowska2019}. 
%All these unusual characteristics make the analysis of astronomical data challenging by standard data analytic tools. 
This is because underlying assumptions  of standard statistical methods do not typically take account of these features.  
For example, even a simple linear regression model requires extra modeling assumptions to account for selection effects and measurement errors for astronomical data analysis \citep{kelly2007}.
Thus, it is important to check these underlying assumptions on a case-by-case basis to ensure sound astronomical data analysis, especially when deploying methods that were developed outside of astronomy.
% of statistical methods
%as is typically appropriate 

%\dvd{From what is below, I think you mean the ``mathematical representation of the mean signal of a physical process''. The Poisson process in intrinsic to many physical processes, but you say this is not part of the physical model.} 
%\vlk{Yes, except that when you say `mean signal' I think you mean a statistical average, whereas an astronomer would call it the `true signal' (well except for the fact that it is predicted from an imperfect model).  If we can find some word other than `mean' the rephrase would work.  `Predicted' perhaps?}
%\vlk{A statistician should look through this and confirm that I haven't mangled or misrepresented the stats concept(s).} 
It is worth pointing out an important distinction in the jargon between the astronomical and statistical literature.  A ``model'' in astrophysics refers to %\dvd\vlk{ok}
a parsimonious mathematical representation of expected (or predicted) signal from a physical process that generates %an 
emission that is eventually detected via telescopes.  This could be the blackbody energy spectrum, or the pulse profile of a pulsar, or the number density of a population of sources in a globular cluster projected onto the sky, etc.  In contrast, a ``model'' in statistics 
%\vlk{I have rephrased this based on David's comment $\rightarrow$}
%.  This includes the physical process of interest, 
is a stochastic representation of the data-generating process that accounts for discrepancies between the astrophysical model and the data. This stochastic representation is indexed in that it is specified up to a set of unknown model parameters that are fit to the data. It reflects systemic adjustments (including observational constraints and instrument effects), selection effects, stochastic components such as Poisson and Gaussian errors, and %everything 
anything else that effects the distribution of the data.
To take a simple example,  the choice of the Poisson$(g(\theta))$ model to represent photon counts forms a statistical model, whereas the astrophysical model stipulates the functional form of $g(\theta)$, e.g., $g(\theta)$ could be a power-law in energy $E$, i.e., $\textrm{norm} \times E^{-\alpha}$,  with model parameters $\theta = \{\textrm{norm},\alpha\}$.
The physical model is designed to describe a physical process without necessarily representing the stochastic aspects of data generation that lead to uncertainty in parameter estimation. Uncertainty quantification, on the other hand, is at the heart of the statistical model which aims to represent data and its variability as fully as possible.
A statistical model also describes the hierarchical connections between the various processes that translate incoming photons to observed electronic signals \citep[e.g.,][]{2001ApJ...548..224V}.

\begin{figure}[t!]
\begin{center}
\includegraphics[width=3.2in]{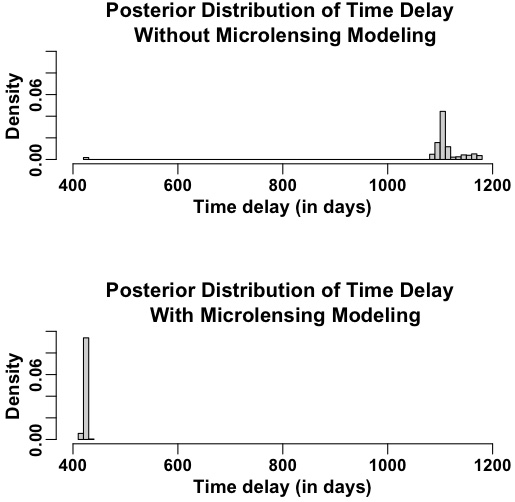}
\caption{The posterior distributions obtained under two different models for the time delay given the same data of two gravitationally lensed light curves of Q0957+561 \citep{2012AAS...21910802H}. The model producing the posterior distribution in the top panel does not account for  the effect of microlensing, producing multiple modes \citep{tak2018ram}, even though the tiny mode near 400 days is of scientific interest. The modes near 1100 days do not appear in the bottom panel where the model additionally accounts for the microlensing effect via polynomial regression \citep{tak2017bayesian}. It turns out that the  modes near 1100 days in the top panel are spurious caused by model mis-specification.
}
\label{fig:timedelay}
\end{center}
\end{figure}

As an illustration of the particular difficulties in handling astronomical data, consider the estimation of the time delay between the multiple images of a strongly gravitationally lensed time-varying source.
In estimating the time delay between gravitationally lensed light curves of Q0957+561 \citep{2012AAS...21910802H},  \cite{tak2018ram} adopt a damped random walk statistical model (also known as a continuous-time auto-regressive model of order one or an Ornstein-Uhlenbeck process) as a data generating process. This model reveals multiple modes in the posterior distribution of the time delay parameter as illustrated  in the top panel of Figure~\ref{fig:timedelay}.  The height of the mode near 400 days is much less than the mode near 1100 days. However, it turns out that the highest mode near 1100 days is spurious, caused by model misspecification. The modes near 1100 days disappear when the astronomical model additionally incorporates polynomial regression to account for the effect of microlensing \citep{tak2017bayesian} that is known to be present in the data \citep{2012AAS...21910802H}; see the bottom panel of Figure~\ref{fig:timedelay}. Consequently,  the mode near 400 days becomes prominent, in agreement with some previous analyses of this quasar \citep{1990AJ....100.1771S, 2008A&A...492..401S}.
% scientific findings about quasar Q0957+561 in the astronomical literature indicate  that the mode near 400 days is of great interest 
%kaisey: The only reason 400 days is of interest is because this was the estimate based on time series analysis of the same quasar, not because of some independent scientific reasons.  This would best be put at the end of the paragraph and restructured to say that "after microlensing is incorporated into the model, the mode near 400 days becomes prominent, in agreement with some previous analyses of this quasar (Schild, Shalyapin)."

%This figure shows the importance of incorporating the story behind the data into a model. 

%the same data used to fit different models
%The maximum likelihood or maximum a posteriori estimate is what the data say in the context of a given model.
This example points out several important aspects in   astronomical data analysis. First, different model fits on the same data  can reveal  completely  different possibilities, e.g., for the time delay of Q0957+561. All of these possibilities are worth proper investigation in the context of available scientific knowledge, in an attempt to determine which are simply the result of  model misspecification and which are new scientific discoveries.  Second, blindly making inference  based on the highest mode of the posterior distribution or likelihood function (or smallest loss function in machine learning methods)
%\dvd{of the posterior distribuiton or likelihodd function} 
can be misleading, as illustrated in the top panel of Figure~\ref{fig:timedelay}.  Thus it is essential to check whether the model captures important characteristics of the data sufficiently well before 
drawing any conclusions.
%\dvd{drawing any conclusions}.
%making an inference.   
Lastly, it is the story embedded in the data that can provide insight for 
improved
%\dvd{improved} 
%extra 
modeling of physical phenomena, such as microlensing.  The better the statistical and astronomical models  reflect the data, the better the quality of what the data reveal
%\dvd{reveal} 
%speak 
to us.

In what follows, we discuss several issues that arise in astronomical data analyses
%related to model mis-specification 
in light of the unique or unusual features of astronomical data. 
We formulate our observations into the following six maxims, each  of which
%\dvd{We formulate our observations into several maxims, each of which}
%and propose a few relevant  maxims. Each maxim 
is in the sprint of George Box's  well-known aphorism ``all models are wrong but some are useful''   \citep{box1987}:
\begin{enumerate}
    \item All data have stories, but some  are mistold.
    \item All assumptions are meant to be helpful, but some can be harmful. 
    \item All prior distributions are informative, even those that are uniform.
    
    \item All models can be given interpretations, but some are more compelling.
    %All models must be interpreted, but some interpretations are more compelling
    
    \item All statistical tests have thresholds, but some are mis-set.
    %\sout{need corrections} .
    %\dvd{It is not clear if we mean the test of the thresholds are miscalibrated.} 
    %(but some thresholds are mis-calibrated?)
    \item All model checks consider variations of the data, but some variants are more relevant than others. 

\end{enumerate}
While we believe that the statement of each of the maxims is new, the ideas that underlie them are not. Rather, the maxims are merely concise statements that we hope capture a sense of the 
%statistical 
reasoning that defines statistics as a discipline and that is the culmination of  the work of
%\dvd{the work of} 
generations of data-facing researchers. Our aim  is to encourage researchers to carefully consider their (possibly implicit) modelling and statistical assumptions and how these assumptions may affect scientific findings. We hope that by keeping the maxims in mind as part of their daily data-analysis routine, researchers will improve the quality of both data modeling and scientific findings in astronomy.

% behind them %stories
\section{All data have stories, but some  are mistold.}
%All data have a story behind them.
%All data are collected, but  some are purposefully.
%data quality
%  (All data are information, but some are mis-information. All data want to speak, but some are speak wrongly. Let the data speak, but not the data lie. All data have a story behind them, but some are mistold)
In this section, we explain several issues in modeling astronomical data, such as sampling mechanisms, selection effects, preprocessing, and calibration, and discuss possible solutions to improve the quality of  astronomical data analysis.

\subsection{Sampling mechanism}
Statisticians typically assume that the data are measurements of a statistical sample that is
%\dvd{measurements of a statistical sample that is} 
representative of the larger class of objects under study. 
For example, we might have a sample of white dwarf stars from the Milky Way Galaxy and measure the metallicity of each or we might have a sample of exoplanets and measure the mass of each.
%\dvd{For example, we might have a sample of white dwarf stars from the Milky Way Galaxy and measure the metallicity of each or we might have a sample of exoplanets and measure the mass of each.}
Formally, statisticians may assume that 
we have obtained a
%\dvd{we have obtained a} 
probability sample from the larger class or population. (In a probability sample, all objects in the population have a known non-zero probability of being included in the sample.) Unfortunately, such a sample is nearly impossible to obtain in astronomy. %One reason is that  
While it is true that measurements of the properties of individual objects have  become more accurate, this accuracy does not translate into a more representative sample of objects. In fact,  none of the so-called all-sky surveys have uniform coverage as they all provide preferential or deeper coverage of specific parts of the sky. For example, Sloan Digital Sky Survey (SDSS) is targeted at the northern celestial hemisphere, the Rubin Observatory has lower coverage in the northern hemisphere, and space-based observatories like TESS and eROSITA have deeper coverage towards the poles.  Likewise, narrowly focused pencil-beam surveys like the {\sl Hubble} Deep Field or {\sl Chandra} Deep Field surveys have varying sensitivity across the field of view due to the detector or telescope responses.

Modeling such data without paying attention to the exact nature of the sampling mechanism and how well they represent the population of interest can result in biased
%how the populations are represented results in incorrect 
inferences \citep[Section 5]{kelly2007}.  Astronomers are generally aware of adverse selection effects introduced by the Eddington or Malmquist biases \citep{1992ApJ...391..494L, refId0}, but we caution that the systematics of any survey or measurement must be carefully considered on a case-by-case basis.

\begin{figure}[t!]
\begin{center}
\includegraphics[width = 3.3in]{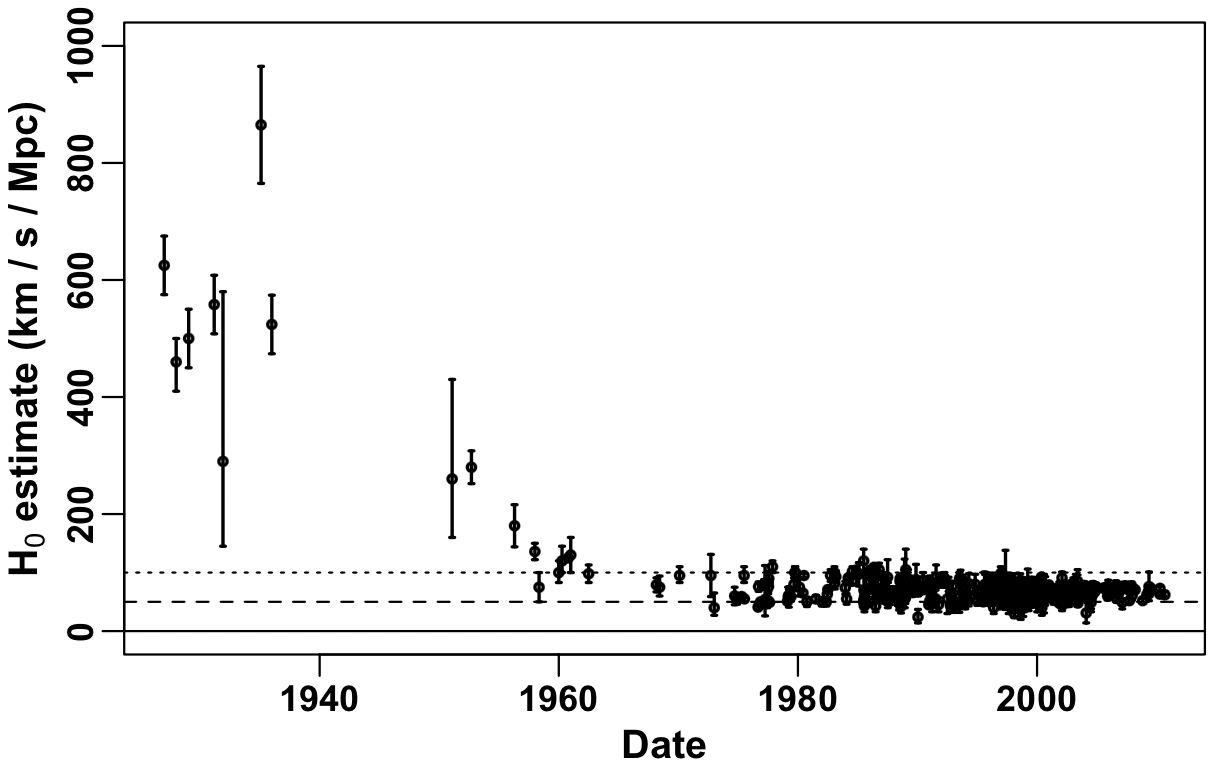}
\caption{A history of Hubble constant estimates made between 1920 and 2008. The dashed horizontal line indicates the $H_0$ estimate at 50 km s$^{-1}$ Mpc$^{-1}$ and the dotted horizontal line represents the estimate at 100 km s$^{-1}$ Mpc$^{-1}$. The half lengths of vertical bars around dots represent $1\sigma$ uncertainty. 
This figure is 
generated %reproduced 
using the 
estimates of $H_0$ compiled  by John P.~Huchra  from the literature %data assembled 
as part of the NASA/HST Key Project on the Extragalactic Distance Scale (\url{https://lweb.cfa.harvard.edu/~dfabricant/huchra/hubble.plot.dat}).}
\label{fig:hubble}
\end{center}
\end{figure}

A well-known example occurs in \cite{1929CoMtW...3...23H}, where systematically high peculiar velocities in the local Galactic neighborhood initially led to a large overestimate of the eponymous Hubble constant, $H_0$. Indeed, the importance of modeling systematic uncertainty is apparent throughout the history of the measurement of $H_0$. 
%the Hubble constant. 
Figure~\ref{fig:hubble} shows that early estimates,
%of  $H_0$
%the Hubble constant 
from the mid-to-late twentieth century, were either around 50~km~s$^{-1}$~Mpc$^{-1}$ denoted by the dashed horizontal line \citep[e.g.,][]{1975ApJ...197..265S} or around 100~km~s$^{-1}$~Mpc$^{-1}$ visualized by a dotted horizontal line \citep[e.g.,][]{1979ApJ...233..433D,1986ApJ...303...19D}.
% \dvd{I think this is all pretty hard to see from Figure 1.}
The half length of each vertical bar around the point estimate represents its 1$\sigma$ uncertainty.

\begin{figure}[t!]
\begin{center}
\includegraphics[width = 3.2in]{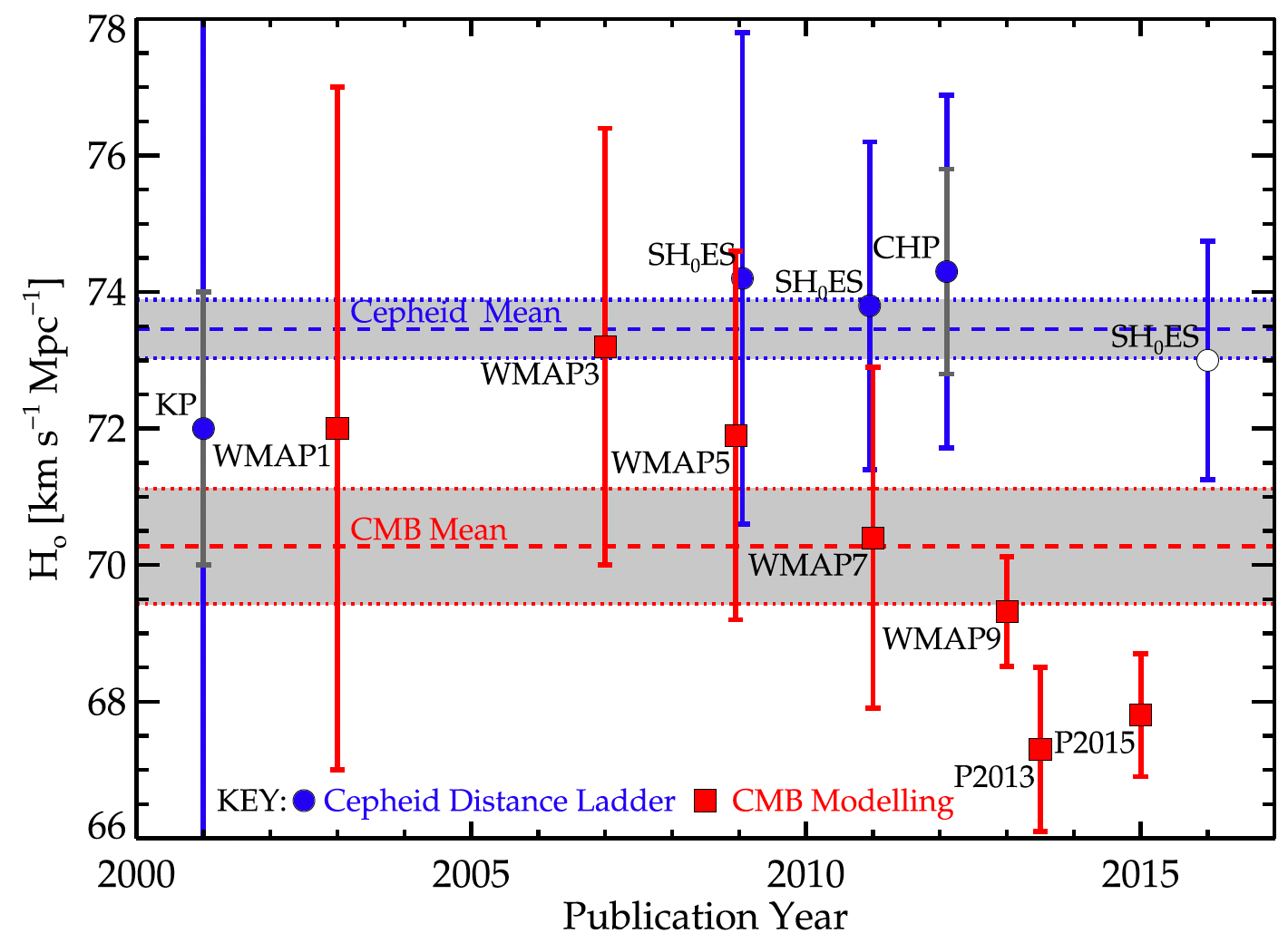}
\caption{The tension between the early- and late-Universe measurements of $H_0$. The estimates in blue are computed by standard candle method calibrated by Cepheid and those in red are obtained by the cosmic microwave background measurements under the standard cosmology ($\Lambda$CDM). This figure is excerpted from Figure~1 of \citet{2016ApJ...832..210B}  by permission of the AAS.}
\label{fig:tension}
\end{center}
\end{figure}

More recently, significant improvements in instrumentation and techniques have led to better understanding of the systematics involved and have narrowed the range of the measured values of $H_0$ further. However, a statistically significant discrepancy
remains among the estimates, raising a question regarding the validity of the standard cosmological model \citep{verde2019tension,2020arXiv200710716E,Riess_2021}. For example, Figure~\ref{fig:tension} (excerpted from Figure~1 of \citet{2016ApJ...832..210B}) illustrates the tension between 
%the Hubble constant 
estimates of $H_0$ from the so-called late-Universe measurements calibrated by the Cepheid distance scale (in blue), and early-Universe measurements obtained by the cosmic microwave background (in red). \cite{2018MNRAS.476.3861F} show that the Bayesian evidence of the standard cosmological model is about seven times smaller than that of an extended cosmological model, that includes an additive deviation  from the standard cosmological model. The corresponding Bayes factor between the two cosmological models is $0.15\pm0.01$ given the Planck 2015 XIII data \citep{2016A&A...594A..13P} and the distance-ladder data of \cite{2016ApJ...826...56R} with extra supernova outliers being considered.

The role of systematics is clear in recent work describing the tension among competing estimates of $H_0$ owing to the extraordinary efforts of the astronomical and cosmological communities to pin down $H_0$. How does a researcher with fewer resources recognize similar effects in their analysis and remedy them?  We posit that this requires an iterative process that implements corrections and appropriately incorporates model complexity in follow-up analyses.  Still, it is important to recognize that any analysis remains vulnerable to imperfect knowledge of the story behind its data.
%\dvd{}

\subsection{Selection  effect}
In addition to non-uniform coverage, astronomical data are often obtained intentionally and purposefully for specific research projects.
%, possibly causing selection bias as well.  
%\textcolor{purple}{(e.g., Budavari \& Szalay, Rots et al)}
When such astronomical data become public through various archives, other researchers may download them and use them as if they were randomly and uniformly selected, possibly unaware of the danger of selection effect in their sample. Likewise, 
when the contents of different surveys are examined together, their individual characteristics can affect the overall interpretation in complex ways.  This is well-appreciated when different catalogs are matched \citep[e.g.,][]{2008ApJ...679..301B, 2017AAS...22915603R, 2017HEAD...1611301R}, but less so when population studies are carried out.
Catalog data are often used as training sets when applying machine learning methods, even though such training sets may not  represent the population of interest well due to the non-uniform coverage and selection effects within the catalog. For example, the Chandra Source Catalog \citep[CSC;][]{2010ApJS..189...37E} provides a selection of fields, each observed for individual scientific reasons. Such ``samples of convenience'' are not probabilistic and not representative of the population, in contrast to flux-limited all-sky surveys like the ROSAT All-Sky Survey \citep[RASS;][]{1993AdSpR..13l.391V,1999A&A...349..389V, 2016A&A...588A.103B}.
% \dvd{really? so there is an importance weight associated with each source in the catalog? \vlk{Not explicitly in the catalog, but all the Chandra fields are only observed because someone thought it was worth observing and a peer review panel agreed.  So, for example, there is $\approx$1.5~Msec of Capella spectral data alone.  Using the CSC for any population analysis literally means we are searching under the streetlamp.}} \dvd{sure: but I wouldn't call that an importance sample. Maybe a ''haphazard non-random sample" or ''convenience  sample"?
%OR:
%For example, the Chandra Source Catalog \citep[CSC;][]{2010ApJS..189...37E} provides a selection of fields, each observed for individual scientific reasons. Such samples of convenience are non-representative of the population, in contrast to ... 
%}, in contrast to 
%\textcolor{purple}{An example of how to deal with these effects is provided by} \cite{revsbech2017} \textcolor{purple}{who} lessen the effect of a biased training set for classifying Type Ia supernovae by augmenting it with multiple pseudo training sets simulated by a Gaussian process. 
%
An example of how to deal with these effects is provided by
\cite{revsbech2017} and \cite{autenrieth2024} who reduce the effect of a biased training set in classifying Type~Ia supernovae via stratification. It forms training sets that are more representative of the corresponding strata within the test set. Similarly, 
\cite{10.1214/16-AOAS1013} propose a non-parametric density estimator for photometric redshift that accounts for selection bias in a non-representative training set by importance reweighting of the training set.  
%vpropose non-parametric conditional density estimators for
%\dvd{Could also mention Peter Freeman's  2017 AoAS paper on photo-Z, dealt with non-representative training data with importance weighting.}

\subsection{Preprocessing}

\begin{figure*}[t!]
\begin{center}
\includegraphics[width=\linewidth]{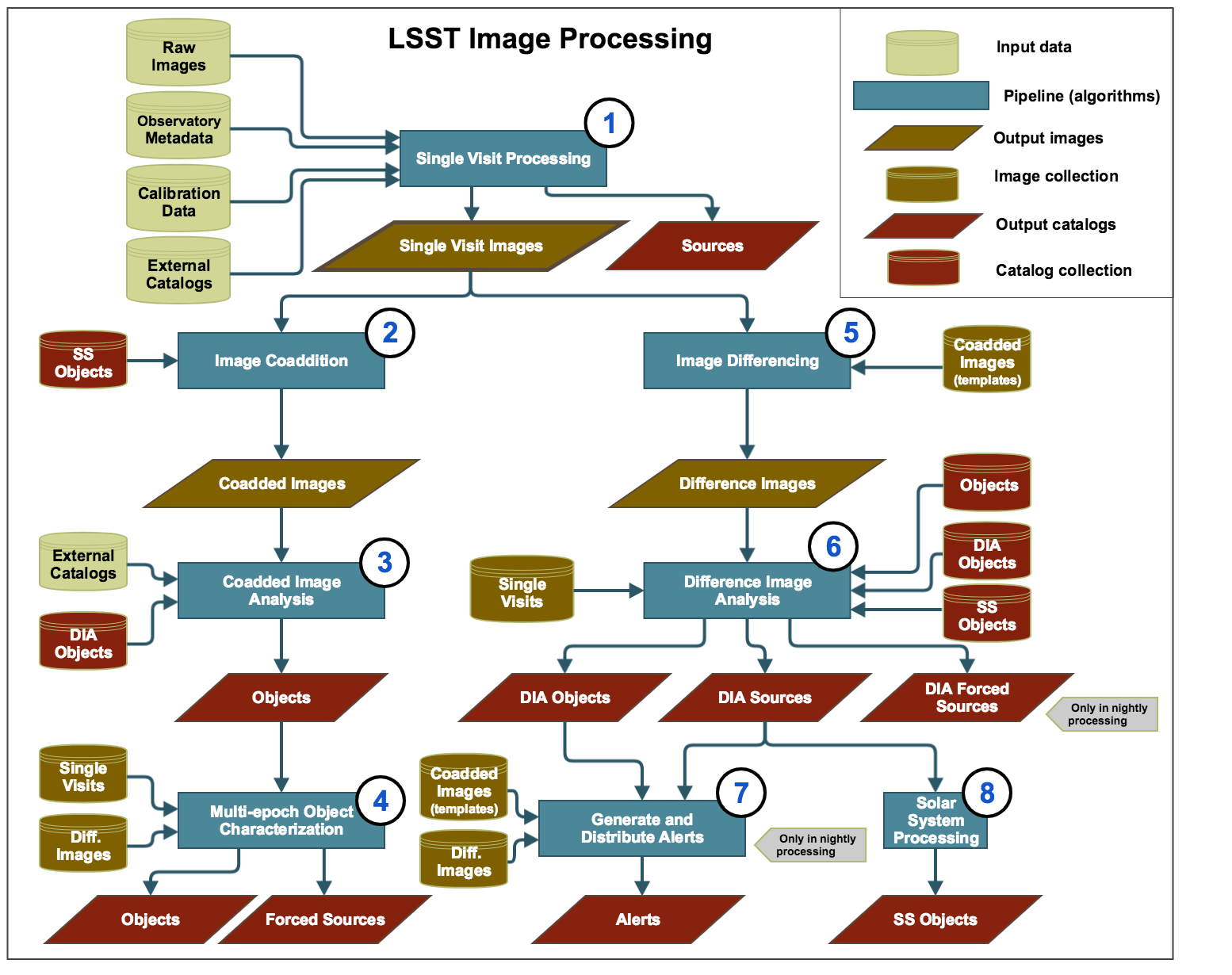}
\caption{A diagram illustrating an example of a data reduction pipeline, excerpted with permission from Figure~2 of the Rubin Observatory LSST Data Products Definitions Document \citep{LSE-163}. Each of the numbered boxes expands into another %(vlk) an equally complicated 
flow diagram. 
While the details in this figure are not important in the current context, it illustrates the complexity of the data pre-processing.
}
\label{fig:preprocessing}
\end{center}
\end{figure*}

Most astronomical data are pre-processed via multi-stage software pipelines specific to a given telescope.
%of telescope. 
As illustrated in Figure~\ref{fig:preprocessing},
in each stage of the pre-processing hierarchy, one astronomer's inference is passed down to be used as an input by the next astronomer. Another inference is then  made with the previous inference being treated as the data.  Unfortunately, this pre-processing 
%However, it 
is often ignored 
%how the data are pre-processed, 
even though the pre-processing steps can reveal evidence of potential systematic errors. For example, in the case of solar flares databases, precision of recorded flare intensities, complex detection/missing characteristics, temperature effects, incompleteness in matched features may all cause systematic errors in the data  \citep{Ryan_2012, Aggarwal_2018}.

%\dvd{Do we need a bit more detail of the solar flares case?} \vlk{REPHRASE SENTENCE ABOVE TO INCLUDE MORE DETAILS ABOUT WHAT THE SYSTEMATICS ARE -- precision of recorded flare intensities, complex detection/missing characteristics, temperature effects, incompleteness in matched features}

%, for instance, \dvd{given the "for example" in the previous sentence, it is not clear what this "for instance" refers to. Maybe drop it? Or "as another example, catalog..."}
As another example, catalog data pre-processing is performed via standard pipelines and assumptions.
%with the pipelines and standard assumption.
This pre-processing procedure generally affects the catalog quality and reliability; outliers may arise if measurements are not performed in a consistent way; different definitions of upper limits may cause an issue of censoring; an incorrectly implemented %a wrong 
pre-processing procedure may introduce systematic error. Thus, it is important to understand how the catalog quantities are derived from the raw data through a chain of pre-processing stages, especially when  different catalogs are  compared or  merged \citep{2008ApJ...679..301B}. 
Whenever possible, the statistical and systematic errors introduced by
%It is also desirable to account for 
the pre-processing procedure should be accounted for within the overall statistical model as much as possible \citep[e.g.,][]{2017AJ....154..132P}.

\subsection{Calibration}

%Finally, c
% \vlk{Include the Fundamental Equation of Observational Astronomy?
% \begin{equation}
%     Y(x,y,channel,time bin) = \int_{Emin}^{Emax} \int_{RA,Dec} \int_{tmin}^{tmax} RMF(E,c;x,y,t) PSF(x,y;E,t) EA(x,y,E;t) f(E,RA,Dec,t)
% \end{equation}}

Calibration is a foundational part of astronomical inference, more so than in any other physical science.  While instrument calibration is indeed used extensively in fields like experimental physics and geophysics, it is of particularly critical importance in astronomy\footnote{Note that the term ``calibration'' is interpreted differently in astronomical, compared to statistical, literature. In astronomy, it refers to the process of characterizing uncertainties and bias corrections induced by instruments, enabling the translation of measured signals into physically meaningful units. In statistical literature, however, calibration generally refers to a process of ``inverse regression'', where measurements of dependent quantities are used to predict corresponding standard measurements, mediated through a known model function. For example, if a functional form $Y=f(X)$ is learned using a training data set, new measurements of a test data set $Y_0$ are used to predict $X_0=f^{-1}(Y_0)$.  This is mainly motivated by instrumental calibration in chemistry, where high-quality ``standard'' measurements $X$ are more time-consuming and expensive than ``test'' measurements $Y$.  The statistical literature includes theories and methods for various linear, non-linear,  multivariate, and dynamic approaches to statistical calibration \citep{osborne1991, kubokawa1994new, brown1995, oman1996exact, rivers2014dynamic, 10.1111/j.2517-6161.1982.tb01209.x}.  Methods are divided into those designed to handle the case where both standard and test measurements have appreciable error, known as comparative calibration \citep{kelly2007, schafer1996likelihood} and methods where the standard measurements are assumed to be perfect or nearly perfect, known as absolute calibration.}.
Astronomical data are for the most part obtained through observations of remote sources, with physical quantities inferred by transforming the observed signals from a detector.  Each telescope or focal plane instrument has its own specific characteristics that affect this transformation, and considerable effort is put into determining these, and tracking changes to them \citep[see, e.g.,][]{2015JATIS...1d7001G,Partridge_2016,PhysRevD.102.122004}.  Ground-based photometric optical astronomy still relies on obtaining regular observations of ``standard stars'' with similar airmass to the target being observed, so even atmospheric variations 
%can be 
must be
%\dvd{must be??}
adjusted for.  High-energy astronomical telescopes construct and store detailed tabular models of the response of a detector to a monochromatic photon\footnote{See OGIP Calibration Memo CAL/GEN/92-002 and addendum \hfil\break(\url{https://heasarc.gsfc.nasa.gov/docs/heasarc/caldb/docs/memos/cal_gen_92_002/cal_gen_92_002.html}, \hfil\break\url{https://heasarc.gsfc.nasa.gov/docs/heasarc/caldb/docs/memos/cal_gen_92_002a/cal_gen_92_002a.html})}, and every mission measures and stores its sensitivity (also called effective area in high-energy astronomy) as a function of photon wavelength. As noted by \citet{villanueva2021}, the details of calibration can have a dramatic impact on the quality of the data.

It is important to understand, however, that the available calibration products are not perfect.  They are the result of measurements carried out in controlled conditions, and thus include measurement errors, as well as systematics that manifest themselves once the instruments are deployed (often in harsh space environments where the chances of radiation damage is high).  Differences in calibration between different instruments must be weighed when different data streams are considered together.  Where available, calibration uncertainty information must be folded in to the analysis \citep{2011ApJ...731..126L,2014ApJ...794...97X}. 
More recently, efforts have been made to derive corrections to effective areas and to source flux estimates based on simultaneous observations of sources with different instruments even in the absence of an absolute reference using multiplicative shrinkage \cite[see, for example,][]{doi:10.1080/01621459.2018.1528978, 2021AJ....162..134M}.
\\
%REMEMBER TO INCLUDE TAYLOR'S PAPER HERE!

There is a common theme in these examples. Knowing the story behind the data allows one to correct for potential model mis-specification, while not knowing the story may leave one oblivious to the same issue. 
Understanding their data, including limitations in the data collection process and potential selection effect, 
enables researchers to
%. With this knowledge, researchers can 
make appropriate corrections themselves. In this way, being attentive to the story  behind their
data enables researchers
%\dvd{data enables researchers}
%and 
to make more reliable inferences regarding their populations and sources
%\dvd{and sources} 
of interest.

\section{All assumptions are meant to be helpful, but some can be harmful.}\label{sec:assumptions}

Popular statistical models were  developed for specific purposes or motivated by particular problems. Some well known models, such as 
Gaussian linear
%\dvd{Gaussian linear}
regression, can be applicable 
in a wide variety of settings
%\dvd{in a wide variety of settings}
%to other cases  
across various disciplines including astronomy and astrophysics with little difficulty. %For 
Another example is  survival analysis which
%\dvd{Another} example \dvd{is}  survival analysis \dvd{which}
is one of the most popular data analyses in bio-medical sciences. In fact, classical survival analysis is  not directly applicable to astronomical data because  censoring in astronomy is due to statistical measurement errors rather than exactly measured failures. Nonetheless, it has  been successfully applied to analyzing left- or right-censored data caused by telescope sensitivity in astronomy \citep{1985ApJ...293..192F, isobe1986}.

However, the 
use
%\dvd{use} 
%blind use 
of well-known models 
without careful consideration of their assumptions
%\dvd{without careful consideration of their assumptions}
must be discouraged because standard statistical models do not account for unusual features of astronomical data or models. \textcolor{black}%{\citep[see, e.g., cautionary constraints on the widely misused $F$-test,][]{protassov2002}}. 
Even the standard linear regression model has underlying Gaussian assumption, while astronomical data may deviate from Gaussianity with outlying observations, low Poisson counts, background subtraction, error propagation, binned data, and/or heavy-tailed and asymmetric distributions.

As another example,
%\dvd{As another example, }
%Moreover, 
standard statistical models, such as linear regression or auto-regressive moving average models,  often assume that %error in
measurement errors are %\dvd{errors are} %is
homoscedastic and unknown. However, astronomical data often come with  one-sigma measurement-error uncertainties that are heteroscedastic and are (assumed to be) completely known \citep{feigelson2021twenty}.
To model these heteroscedastic measurement-error uncertainties,
community efforts have been made in various fields of astronomy. Introducing 
errors in measurement
to standard models is a popular idea in statistics \citep{fuller1987}. The technique was later tailored to heteroscedastic measurement errors in astronomical data in the contexts of (but not limited to) linear regression \citep{akritas1996linear, kelly2007, andreon2013measurement, sereno2016bayesian}, damped random walk process \citep{Kelly_2009, Hu_2020},  continuous-time ARMA($p, q$) process \citep{kelly2014flexible, 2023ApJ...950...37M}, and astronomical object classification \citep{2011ApJ...729..141B, 2022AJ....164....6S}.

%are compatible with most standard models 
Checking the assumptions of popular models is often facilitated by well-defined model checking procedures.  For example, checking model assumptions via residual analysis is common in regression because it can provide insight into possible improvement for the current model fit. \cite{tanaka1995} improve a poor continuum fit of spectral data via subsequent residual analysis; \cite{bulbul2014} and \cite{reeves2009}  detect emission lines and absorption lines via residuals;  \cite{mandel2017} compare conventional and proposed  models for the color-magnitude relation of Type~Ia supernova  by checking their Hubble residuals to see which model is better supported by the data.

Residual analysis often provides hints that can be used to improve model assumptions.
For instance, when a model for light curves 
such as a damped random-walk model
%\dvd{such as a damped random-walk model} 
relies on a Gaussian measurement error assumption %, such as a damped random-walk model 
\citep{Kelly_2009}, a residual analysis may reveal some evidence against the Gaussian assumption in the presence of outliers. In an effort to improve residual analysis,
\citet{tak2019robust} and \citet{2022MNRAS.516.5874W}  derive a heavy-tailed version of the damped random-walk model that 
is still able to constrain
%contains 
%\dvd{is} still \dvd{able to constrain} %contains 
the same model parameters %of interest 
in a  robust manner.  

Besides standard model checking procedures, investigating the fitted model in light of the knowledge of domain science is also crucial as it can reveal evidence of potential model misspecification. For instance, the sensitivity or dependence of model fits on the
%when model fits noticeably vary according to} 
starting values of optimizers or Monte Carlo samplers is not necessarily an indication of a numerical problem. It could instead point to a multi-modal outcome on a non-convex surface of the parameter space, providing several distinct model fits at different modes. Considering that a model describes a data generation process, it also implies that distinct sets of parameter values for a given model could have generated the same observed data, even though each set  may not be equally likely to have generated the observed data.

Alternatively, a multi-modal likelihood function may indicate that the model is misspecified or is not elaborate enough to describe the data, either of which can lead to an unrelaible fit. Mode-based estimates,  such as maximum likelihood estimates and posterior modes, aim to compute the parameter values corresponding to the particular model within the posited class that best matches the data (under a criteria determined, e.g., by the likelihood or posterior). Even the best match within the posited class of models, however, may not be very good if the class is not sufficiently rich. Using the fully Bayesian posterior distribution with a misspecified model can also lead to unreliable results, as emphasized in Figure~\ref{fig:timedelay}, where multimodality disappears when %after 
we additionally model microlensing. %about microlensing. 
We emphasize that a well specified model is key to any model-based method, and thus it is critical for researchers to  check the fit of their posited model  in light of domain science knowledge, instead of blindly proceeding with the highest mode or other computed summary as the best model fit.

Another popular estimation tool in astronomy is $\chi^2$-minimization that is built on a Gaussian approximation for the measurement error. 
For example, when the data are binned\footnote{We consider the case where the data are intrinsically binned. When this is not the case, \cite{feigelson2012} suggest avoiding issues of arbitrary binning by using cumulative distribution function for  maximum likelihood estimation.} Poisson counts, %it is 
a Gaussian approximation to the Poisson counts
is required for $\chi^2$-minimization \citep{cash1979, 2009ApJ...693..822H, 2020JApSt..47.2044B}.  Thus, it is important to understand the limitations that might affect the validity or accuracy of this approximation. The method is often misused in the context of astronomical data analysis, for example, when the estimated variance of the approximate Gaussian distribution is quite different from that of  the observed (or average) count, which contradicts the validity of Gaussian approximation to Poisson counts \citep[Chapter 7.4]{feigelson_babu_2012}. The approximation itself can be quite misleading when the underlying Poisson assumption is not appropriate, e.g., when the count data are overdispersed. The approximation becomes less accurate when counts of some bins  are small. In this case, merging adjacent small bins  is one way to improve the accuracy of the approximation while sacrificing the resolution of the data \citep{greenwood1996}. 

Directly building a Poisson model for the counts
%Building a model directly on Poisson counts 
without using a Gaussian approximation is another possibility.  (This has not always been well recognized among astronomers \citep{hilbe2014}.)  For example, \cite{cash1979} proposes the so-called $C$-minimization technique, which is operationally the same as finding the maximum likelihood estimate under a Poisson likelihood function. \cite{2020JApSt..47.2044B} demonstrates that $C$-minimization outperforms $\chi^2$-minimization with low-count data because the former does not involve a Gaussian approximation. Also, \citet{2012ApJ...752...55K} adopt Bayesian hierarchical modeling in fitting spectral energy distributions on flux data, instead of using $\chi^2$-minimization. Hierarchical modeling also provides a mechanism for effectively handling overdispersed data \citep{gelman2013bayesian, JSSv078i05}.
Another benefit that direct likelihood-based modelling has over $\chi^2$-minimization is that it facilitates the use of information criteria for model selection, such as the Bayesian Information Criterion \citep{kass1995}, which depend directly on the likelihood.

The central limit theorem is the basis for a G\textbf{}aussian approximation to Poisson counts and generally plays an important role in statistical inference.
It stipulates that the distribution of the average of independent observations becomes more Gaussian as the number of observations approaches infinity. The theorem is the basis of many asymptotic results such as the asymptotic normality of maximum likelihood estimates, %and maximum a posteriori estimates, as well as 
the asymptotic $\chi^2$ distribution of the likelihood ratio test statistic via Wilk's theorem \citep{wilks1938large},
and the ``projection method'' for  computing error bars \citep{avni}.

To be confident of the applicability of asymptotic results, researchers must check two things: 
%\begin{enumerate}
    %\item 
     the assumptions required for the results are met  %meaning that with a sufficiently large dataset, the asymptotic results serve as a good approximation, 
    and  
    %\item 
     the data set they are analyzing is sufficiently large. 
First, all asymptotic results depend critically on their own sets of mathematical assumptions known as regularity conditions. Even with an arbitrarily large data set, the central limit theorem 
itself fails, for example,  if the expected value of the square of the averaged observations is not finite (e.g., when averaging ratios) or if the number of parameters increases sufficiently quickly compared to the sample size \citep[e.g., as with instrument calibration][]{chen2019}).
The likelihood ratio test that compares the statistical evidence for two posited models is another example where the regularity conditions play a key role. This is because Wilk's theorem only provides the asymptotic distribution of the likelihood ratio test statistic if, among other conditions,  the models being compared are nested (i.e., one model is a special case of the other) and the simpler model is not on the boundary of the parameter space describing the more complex model. \citet{protassov2002} show that the latter condition fails when testing for an added spectral emission line because 
an {\it emission} line by definition cannot have a negative normalization but the normalization is zero in the simpler model (with no line).

Second, even if their regularity conditions are met, asymptotic
statistical methods % based on asymptotic results 
are only reliable with sufficiently large data sets. %that are  sufficiently large.} 
A likelihood function that exhibits multiple significant modes, for example, may be evidence that either the model is misspecified (and a regularity condition is not met) or the data set is not sufficiently large for the asymptotic Gaussian properties of the likelihood to have ``kicked in''.
In practice, it can be difficult to know whether a data set is large enough. Generally, the more fitted parameters, the more data that are required. Goodness-of-fit tests are a particular challenge when the data size is small, because, in effect they compare the posited model with a fully flexible model, i.e., a model with a large number of fitted parameters. The $C$-statistic\footnote{The $C$-statistic is sometimes called the Cash statistic in recognition of \citet{cash1979} and is defined in different ways by different authors. For example, it is often defined either as $-2$ times the Poisson log-likelihood function \citep[up to an additive constant, e.g., Eq.~(5) of][]{cash1979} or as $-2$ times the log of the likelihood ratio comparing a specific Poisson model with a fully saturated Poisson model \citep[e.g., Eq.~(1) of][]{2009ApJ...693..822H}.
Both definitions are equivalent up to a constant adjustment. The latter is a particular instance of the likelihood ratio test statistic, but with only the alternative likelihood evaluated at its maximum likelihood estimate.  We use term $C$-statistic for the same likelihood ratio test statistic, but follow the standard statistical convention with both likelihoods being evaluated at their respective maximizers; see Section~\ref{maxim6} for details.}, %(or Cash Statistic), 
for example, is often used for goodness of fit tests in high-energy spectral analysis \citep{kaastra2017use}. If the $C$-statistic is applied to high-resolution data with many narrow bins, %the comparison between observed counts and expected counts under the model is made in each each of a large number of bins (e.g., according to the natural resolution of the data), 
its asymptotic $\chi^2$ distribution %of the C-statistic 
is only guaranteed %by Wilk's Theorem 
if the expected counts are large in {\it all} bins. %This is not typically the situation in practice. 
The alternative is to work with fewer larger bins, %with higher expected counts, 
but this sacrifices the resolution of the data.

When there are insufficient data for asymptotic results, %In practice, we may not be able to gather large enough datasets for an asymptotic approximation to be sufficiently accurate. Luckily both 
either Bayesian procedures or bootstrap-based methods can be used with small data sets. Unfortunately, both are  computational more costly than their asymptotic frequentist counterparts. Higher-order asymptotics, which retain more terms in their functional expansions, sometimes can show advantages in such scenarios. For example, \cite{li2024cstat} obtain a computationally efficient and statistically precise procedure for goodness-of-fit tests based on the $C$-statistic and higher-order asymptotics. The method only involves calculation of moments and works even in low-count settings where the Wilk's theorem ($\chi^2$ asymptotics on likelihood ratio tests) does not apply.

%~\citep[a.k.a. Cash statistic, see][]{kaastra2017use}

\section{All prior distributions are informative, even those that are uniform.}
% always convenient, influential
%kaisey: Should this be generalised to so-called 'non-informative priors'  (for example Jeffreys priors)

Although Bayesian analysis has become popular in astronomy \citep{pierson2013, 2023arXiv230204703E}, it is difficult to find an article that conducts a Bayesian analysis without using uniform priors (often uniform on the logarithmic scale). 
One possible explanation for this popularity 
may actually be
a misunderstanding, 
namely
a perception
%\dvd{a perception} 
that the interpretation of Bayesian inference is more straightforward than that of frequentist inference. For example, one might think that a credible interval is a direct statement about the unknown parameters given the data, while confidence intervals need to be interpreted under a hypothetical repeated sampling scenario of the data.  However, it is often forgotten that the interpretation of credible intervals hinges on the interpretation of the prior distribution, which can be philosophically as subtle as frequentist's repeated sampling scenarios because prior distributions are chosen by researchers.

For instance, uniform priors are often assumed on a logarithmic scale, that is, $\log(X)\sim \text{Unif}(a, b)$, where $a$ and $b$ are real-valued. One may be tempted to interpret the resulting credible interval as if a non-informative prior were used on the original scale, i.e., on $X$. A uniform prior on $\log(X)$, however, can be very informative indeed on $X$, since it corresponds to a power-law prior distribution on $X$ ($d\log(x)=dx/x$)  that  puts substantial probability mass near the lower bound, $e^a$. Thus, the  posterior distribution depends strongly on whether a uniform prior is assumed on the original or logarithmic scale, and the resulting credible interval needs to be interpreted accordingly. In general, it is a mathematical fact that any prior distribution carries information to be interpreted, as it must specify how likely one state is relative to another; see Section~7 of \cite{craiu2023six} for more discussion.

In some sense, uniform prior distributions and other so-called ``non-informative'' prior distributions
%might 
have lessened the burden of subjectivity and prior interpretation for astronomers, making the likelihood (i.e., the data) a dominant source of the posterior variability. In some cases, they also enable a Bayesian inference to be conducted relatively easily for researchers who prefer a Bayesian approach, e.g., to handle nuisance parameters or for uncertainty quantification \citep{e19100555},  even when the maximum likelihood estimate is nearly identical to the maximum a posteriori estimate. Moreover,
uniform prior distributions provide researchers a way to %provide a way to 
incorporate scientific knowledge via their boundaries. 
In most articles,  the bounds of uniform priors are clarified to avoid potential posterior impropriety \citep{tak2018how}.  

%more about 

\begin{figure}[t!]
\begin{center}
\includegraphics[width=3.2in]{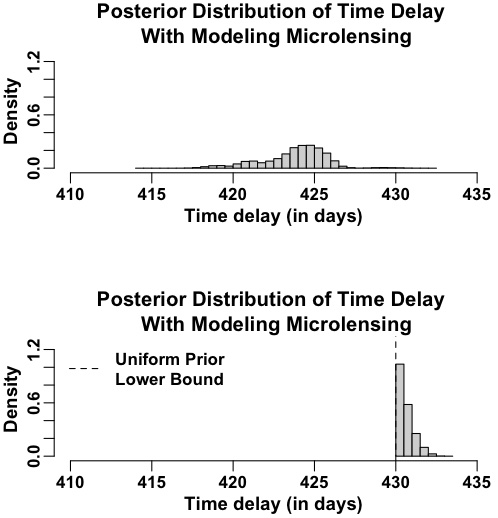}
\caption{The posterior distribution of the time delay under the microlensing model. The top panel is based on the uniform($-1178.939, 1178.939)$ prior distribution for the time delay parameter, while the bottom panel is based on the uniform(430, 1178.939) prior distribution. In the bottom panel, the posterior mass accumulates near the lower bound, which may be evidence of %model 
mis-specification of the prior distribution.
%\dvd{of the prior distribution}.
}
\label{fig:timedelay2}
\end{center}
\end{figure}

Even bounded uniform prior distributions, however, must be used with care because the bounds are hard bounds that completely exclude
a portion
%\dvd{a portion} 
%some regions
of the parameter space. An issue may occur if the bounds partially or completely exclude important regions of the parameter space a priori. For example, the top panel of Figure~\ref{fig:timedelay2} magnifies the posterior distribution of the time delay under the microlensing model, previously shown in the second panel of Figure~\ref{fig:timedelay}. The prior distribution for the time delay adopted in \cite{tak2017bayesian} is the uniform prior between $-1178.939$ and $1178.939$ days, reflecting the widest range of observation times in 
the analyzed
%\dvd{the analyzed} 
light curves. As an illustration, let us set up a different lower bound of this uniform prior at 430 days,  excluding the modal location at around 425 days. The bottom panel of Figure~\ref{fig:timedelay2} exhibits the resulting posterior distribution of the time delay. The posterior mass accumulates near the lower bound, as if the posterior mass in the top panel were pushed from the left to the lower bound. This is what happens when the hard bound of a uniform prior 
zeros out
%\dvd{zeros out}
%deters 
the likelihood 
beyond
%\dvd{beyond} %to jump over 
the bound. The likelihood cannot overcome this hard bound, regardless of 
how large the data set is;
%\dvd{how large the} data \dvd{set is;}
%size; 
even one trillion observations 
would not allow posterior probability beyond the bound. 
(\citet{lindley1985} warns against assigning a probability of zero to events that are not logically impossible in what is often referred to as Cromwell's rule\footnote{The reference to Oliver Cromwell refers to a quotation of his: ``I beseech you [to] think it possible that you may be mistaken''.}.) 
%mentioned in \cite{lindley1985}). 
%\dvd{would not allow posterior probability beyond the bound.} 
%cannot make it cross the bound. 

\begin{figure}[t!]
\begin{center}
\includegraphics[width=3.4in]{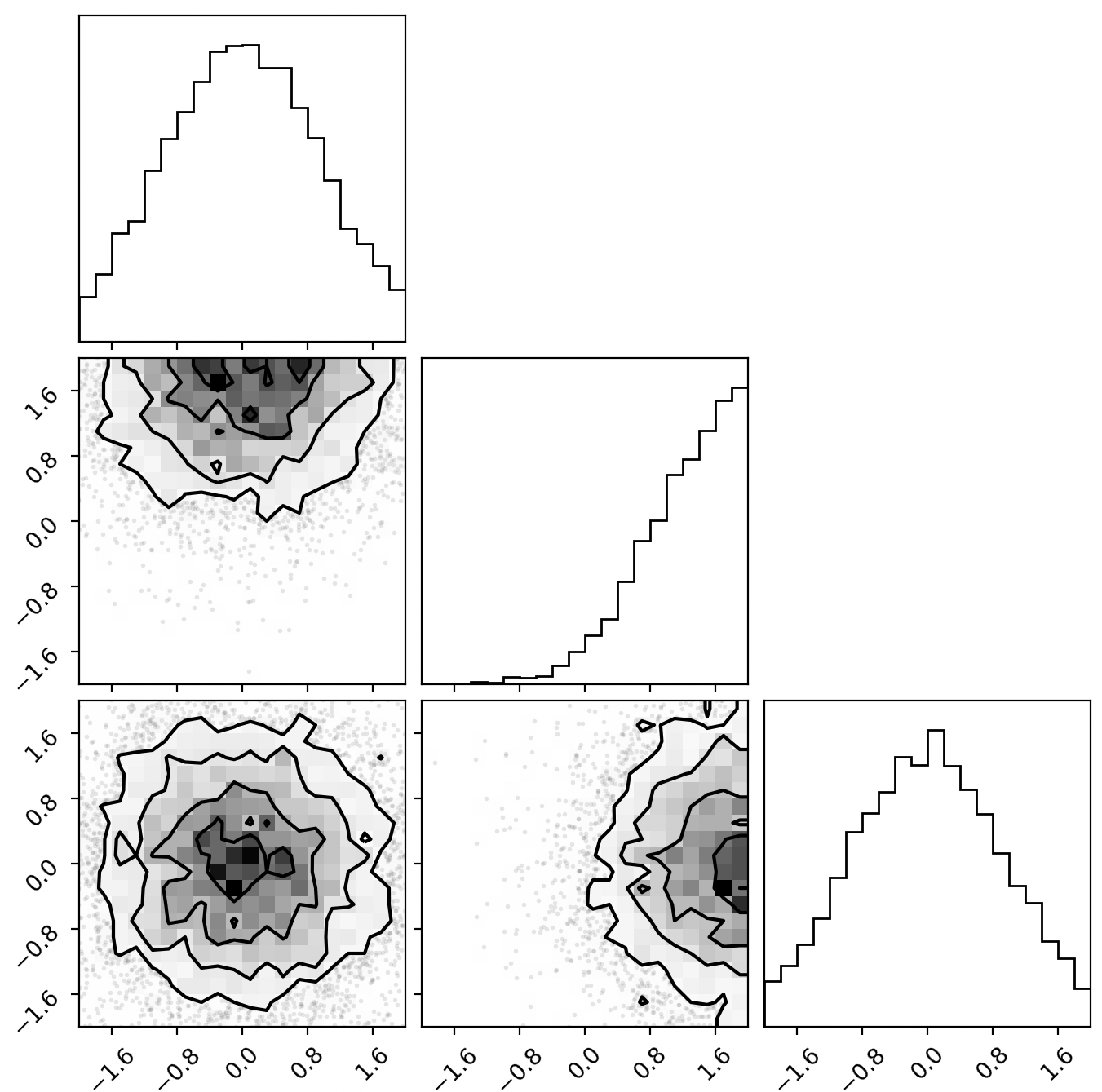}
\caption{A simulation study where the posterior mass accumulates near the boundary of a uniform prior distribution. The uniform prior distribution of the second parameter is bounded between $-0.2$ and 0.2, and the posterior mass clearly concentrates on the upper bound. In general, when a uniform prior distribution is used, it is important to check the sensitivity of the resulting posterior distribution to the choice of hard bound on the prior,  unless there is a specific scientific justification for the choice of bound.}
\label{fig:bound}
\end{center}
\end{figure}

Substantial posterior probability that is accumulated near the (hard) bounds of a uniform prior distribution may be evidence 
of mis-specification of the bounded prior distribution.
In the astronomical literature, it is not difficult to find examples with substantial posterior mass near the  bounds of uniform priors that are set by researchers. 
This problem can often be identified by inspecting corner plots (pairwise scatter plots with marginal histograms) in published articles, at least when these plots are provided by the authors. A simulated example similar to one we found in the literature survey\footnote{For instance, a search for articles that include the word `Bayesian', published in MNRAS in June 2024 yields 58 articles, of which 17 displayed corner plots; 7 of these plots showed boundary issues.} is shown in Figure~\ref{fig:bound}. When a researchers identifies a boundary issue of this sort, it is critical that they carefully investigate the sensitivity of their results to the bounds of their uniform prior, paying particular attention to the robustness of their scientific conclusions to the choice of bounds.
 Where there is a natural bound, e.g., where a parameter such as a mass or age must be non-negative or positive,
 we do not consider the 
 accumulation of posterior mass near this natural bound to be
 an issue.
 Therefore, unless there is a strong scientific justification, it is always better to set uniform bounds wide enough not to influence the likelihood.

% The figure is excerpted from a published article, but we do not specifically cite the paper because we use this only as an illustration of a common issue in the astronomical literature.

 %\dvd{I don't like the over anthropomorphic descriptions of data and models and likelihoods speaking. I think it is strained and a distraction.}

Besides the boundary issues of uniform priors,  a blind use of jointly uniform prior distributions can become a highly informative choice, despite its seemingly non-informative nature \citep[p223]{gelman1996}. For example, when model parameters are constrained such as being in an increasing order (in astronomy, for example, unknown breaking points in multiply broken power laws), a jointly uniform prior on the parameters 
asymptotically
%\dvd{asymptotically} 
dominates the likelihood function as the number of such  model parameters increases. \cite{e19100555} provide more examples where uniform prior distributions can result in  inferences that do not make sense.  A jointly improper uniform prior can also be problematic  in high dimensions, even though it results in a proper posterior distribution; Section 4.2 of \cite{e19100555} discusses a similar problem that arises when independent Gaussian prior distributions are used in high-dimensional parameter spaces.

\section{All models can be given interpretations, but some are more compelling.}
%All models must be interpreted, but some interpretations are more compelling
%All models are subject to interpretation, but some interpretations are less contrived
%All models have interpretations, but some are .

Understanding how the statistical/mathematical interpretation of an empirical data analysis should impact our understanding of astrophysical processes can be challenging. We suggest that it is often best to start with the physics and then consider whether the empirical findings make sense in terms of the physics and/or how we can make sense of them.

\cite{Kelly_2009}, for instance, carefully investigated a sample of quasar light curves, relating two model parameters of a damped random walk process to physical properties of a quasar. They show that the timescale (short-term variability) of the fitted process is positively (negatively) correlated to both black hole mass and luminosity.   This empirical evidence on the relationships between the parameters of the mathematical model and astrophysical properties has since been intensively investigated and is supported by many astronomers \citep{macleod2010modeling, kozlowski2010quantifying, kim2012, andrae2013}.

For more elaborate model interpretation, the community has also investigated when this interpretation does not hold. \cite{mushotzky2011kepler} and \cite{zu2013is} show that the damped random-walk process is not suitable for explaining the stochastic variability of quasars when the source variability is on a very short timescale. Also, \cite{graham2014a}, and \cite{kasliwal2015are} warn that the process is too simple to explain all types of stochastic variability of quasars. When the underlying true model is not the damped random walk process,  \cite{kozlowski2016deg}  points out that the association between the model parameters and physical properties can be misleading as  timescale estimates become biased.

%, that enables modeling various types of stochastic variability  can be modeled  
This productive discussion has motivated astronomers to consider the more general and flexible class of models known as continuous-time auto-regressive moving average processes \citep{kelly2014flexible}. This class encompasses a wide variety of stochastic processes and requires users to select 
%the values of parameters specifying 
the orders of both the auto-regressive and the moving average components of the model. The Akaike information criterion \citep{akaike1974new} is a popular model selection criteria that can be used to select these orders \citep{kelly2014flexible, 2019AJ....158...57C}, but may not exhibit well calibrated statistical properties \citep[e.g.,][]{sutherland2023practical}. \cite{2023ApJ...950...37M} illustrate an alternative strategy based on the Bayesian evidence. This is well-grounded theoretically, but results should be checked for  sensitive to the choice of prior distribution. Both strategies aim to identify an appropriate level of model complexity and thus avoid both
under- and over-fitting.
%This class requires model selection in practice as it encompasses wide variety of stochastic processes that vary according to the order of the auto-regressive model and that of the moving average model. For example, Akaike information criterion \citep{akaike1974new} is one of the most popular model selection criteria used for this class \citep{kelly2014flexible, 2019AJ....158...57C}, and \cite{2023ApJ...950...37M} use Bayesian model selection for this class based on the Bayesian evidence. These model selection approaches enable researches to identify the most appropriate model for the given data, effectively preventing 

Substantial community effort has also been devoted to investigating the applicability and physical interpretation of the resulting power spectral densities via empirical evidence \citep{2019PASP..131f3001M, 2022ApJ...936..132Y}. This class of models is promising as it can be extended to model long-memory auto-correlations \citep{10.3150/bj/1165269152, marquardt2007generating}, although its limited applicability to stationary time series data  remains.

This collective community effort  is the key to  building time-tested models with widely accepted   astrophysical interpretations  as it is crucial to demonstrate the empirical evidence and  effectively warn  of cases where the interpretation of the model parameters is fallible.

\bigskip
\section{All statistical tests have thresholds, but some are mis-set.}
%All hypothesis tests come with rejection thresholds, but some need more corrections 
%\\ 
%\vlk{All tests come with thresholds, but some need corrections.}
%\\
%\dvd{All statistical tests have thresholds, but some must be corrected.}

Hypothesis testing compares two models 
for the same data; the two models are called
%\dvd{for the same date; the two models are called} 
the null and the alternative hypotheses.
%given the same data. 
In the standard frequentist setup, the researcher specifies a test statistic with known (or approximately known) distribution under the null hypothesis. Inconsistency between the observed value of test statistic (computed using the research data) and this null distribution is viewed as evidence that the data is unlikely to have arisen under the null model and thus as evidence in favor of the alternative hypothesis. Inconsistency is typically measured with a $p$-value, the probability that a value of test statistic as extreme or more extreme than the observed value would arise under the null distribution. The principled usage of this paradigm
%Its principled usage 
is crucial in various scientific fields because it is the procedure that provides data-driven evidence for a scientific discovery or anomaly against well-established 
theories.
%\dvd{theories}
%knowledge. 
Unlike 
in
%\dvd{in} 
biomedical research, hypothesis testing in astronomy is less likely to suffer from 
common
%\dvd{common}
%popular 
issues regarding $p$-values, such as a blind usage of ``$p$-value $<$ 0.05'' or $p$-hacking, for example, collecting more data until the $p$-value becomes smaller than 0.05 \citep{wasserstein2016}. This is partly because astronomers typically use more conservative thresholds 
for establishing
%\dvd{for establishing} 
statistical significance, such as a 3$\sigma$ level \citep{2023PhRvD.108l3007V}, making 
significance
%\dvd{significance}
%it
more difficult to achieve by simple data manipulation\footnote{Early usage in astronomy tended to use the $2\sigma$ (5\%) thresholds \citep[see discussion in][]{1979QJRAS..20..138W}.  Higher thresholds have also been adopted; e.g., the processing pipeline for the {\sl Einstein} X-ray observatory used $4-5\sigma$ thresholds \citep{1984SAOSR.393.....H}.}. This 3$\sigma$ threshold corresponds to type I error rate of $\alpha$ equal to $0.0027$ (that is, the probability of rejecting the null when it is correct) for a two-sided test when a test statistic is distributed as $N(0, 1)$ under $H_0$.
%, typically denoted by $\alpha$ in the statistics literature,
%, and the observed value of the test statistic is 3

A common application of significance testing in astronomy has been for the purpose of source detection.  A ``$3\sigma$'' detection usually implies that the ratio of the estimated flux to its error is $\ge3$.  It is worth noting that this ties ``detectability'' to flux estimation.  Newer methods like {\sl CIAO}/{\tt wavdetect} \citep{2002ApJS..138..185F} explicitly separate detection from flux estimation, carrying out the former via a $p$-value threshold set based on the estimated background alone.  Such methods have more power and improve the sensitivity of the detectors, enabling the detection of weaker sources\footnote{Consider a case where the {expected} background under the source is precisely estimated to be 0.9 counts. {The detection threshold is set to the count where}
%Then, the counts threshold that is set for 
the cumulative tail probability of the Poisson distribution {under the background only model} %to  
drops below the $3\sigma$ significance, {i.e, the detection threshold is 5 counts. 
The probability of observing 5 counts or more if there were no source is} 
%is 4 counts.  %Thus, if 4 counts are observed, the probability of its occurrence is 
0.00234, less than the adopted threshold of $p=0.0027$. {Thus if a count of five or more were observed} %and 
the source would be declared detected.  In contrast, the signal-to-noise ratio based on the estimated background-subtracted source strength and Gaussian error propagation approximations \citep{1986ApJ...303..336G} can only exceed 3 when $>15$ counts are observed; the source would be declared detected if and only if its flux can be estimated with a sufficiently small uncertainty.}.  Using only the background also allows a statistically well-defined definition of an {\sl upper limit} to the source flux, which can be set as the source intensity that would be detectable with a specified power \citep{2010ApJ...719..900K}. 

% pay attention to 

%that sometimes lacks in the hypothesis testing procedures in astronomy is 
One aspect that astronomers must keep in mind when interpreting  significance levels of multiple test statistics, however, is 
how to
%\dvd{how to} 
control the family-wise error rate (FWER). The FWER is defined as the probability of committing at least one type I error (false-positive) among $m$ hypothesis tests, and is smaller than or equal to $1-(1-\alpha_{\text{ind}})^m$. The notation $\alpha_{\text{ind}}$ denotes the common type I error rate used for 
each of the $m$ individual
%\dvd{each of the $m$ individual} 
hypothesis tests. A good example to illustrate the FWER can be found in \cite{2016PhRvL.116f1102A}, where the detection of the first gravitational wave  is based on the 4.6$\sigma$ and 5.1$\sigma$ significance levels of two test statistics, respectively. Clearly, each of the reported significance levels is greater than the 3$\sigma$ threshold. However, naively comparing each reported significance level with the 3$\sigma$ threshold is equivalent to 
maintaining an
%\dvd{maintaining an} 
%controlling the 
FWER 
that is
%\dvd{that is} 
less than or equal to 0.0054 ($=1-(1-0.0027)^2$). That is, the probability of committing at least one type~I error in the two tests for the gravitational wave detection is actually twice 
as large as
%\dvd{as large as} 
%larger than 
the individual type I error rate. To 
ensure
%\dvd{ensure}
the FWER 
is less than
%\dvd{is less than}
%to be smaller than 
or equal to 0.0027, as intended, the popular Bonferroni correction \citep{amstrong2014} sets the individual type I error rate to be 0.00135 ($=0.0027/2$), which requires comparing each of the reported significance levels with a 3.2$\sigma$ threshold, not $3\sigma$.

One possible issue with the Bonferroni method is that it is rarely possible to reject a null hypothesis when the number of hypothesis tests $m$ is large. For example, to 
ensure that
%\dvd{ensure that}
%control 
the FWER 
is less than
%\dvd{is less than}
%to be smaller than 
or equal to 0.0027 among 1,000 hypothesis tests, the individual type I error rate must be $2.7\times 10^{-6}$, which is a threshold that may be
%is very 
difficult for individual $p$-value to achieve. This can lead %leads 
to almost no rejection among the 1,000 hypothesis tests. %One way to alleviate this issue of FWER in controlling false-positives in multiple hypothesis tests is to
As such, the Bonferroni correction can be too conservative for certain testing scenarios in astronomy, for example, for compiling astronomical catalogs.
% for many scientific purposes (citation?)

An alternative is to instead control the false discovery rate \citep[FDR,][]{1995benjamini, benjamini2010}. Unlike the FWER, the FDR 
ensures that
%\dvd{ensures that} 
%controls 
the expected proportion of false-positives among all of the false- and true-positives 
is less
%\dvd{is less}
%to be smaller 
than or equal to a preset value. That is, controlling the FDR 
to be less than or equal to
%\dvd{to be less than}
%below
0.0027 means that the proportion of false-positives (false discoveries) among all of the rejected null 
hypotheses
%\dvd{hypotheses}
(discoveries) is less than or equal to 0.27\%.

In practice, controlling the FDR results in a different procedure for rejecting the null hypothesis. For example, 
suppose we wish to ensure that the FDR is less
%\dvd{supppose we wish to ensure that the FDT is less}
%let's say we want to control the FDR to be smaller 
than or equal to 0.0027 among 1000  hypothesis tests. We first sort the 1000 $p$-values in an increasing order, $p_1, p_2, \ldots, p_{1000}$, where $p_1$ is the smallest $p$-value and $p_{1000}$ is the largest. Next, we find the largest  index such that $p_i< (0.0027\times i) / 1000$. If this largest index were 55, for instance, then we 
would
%\dvd{would}
reject the null %hypothesis in
hypotheses associated with
%\dvd{hypotheses associated with}
the first 55 tests. Note that the Bonferroni correction in this case is to reject the null hypothesis in test $i$ if $p_i< 0.0027 / 1000$.  Consequently, the FDR results in more rejections (possibly more false-positives,
but also more true-positives)
%\dvd{, but also more true-positives}) 
than the FWER. This is a useful feature 
of FDR
%\dvd{of FDR} 
when the number of tests $m$ is large. Moreover, controlling FDR is known to be more powerful than  controlling FWER, while the former comes with higher type I error rate (more false-positives) \citep{shaffer1995}. A receiver operating characteristic (ROC) curve can be useful for investigating the balance between the false-positive and true-positive rates  obtained using different values of the FDR (or FWER) threshold.

In \cite{2016PhRvL.116f1102A} for the detection of the first gravitational wave, the two $p$-values corresponding to the reported 4.6$\sigma$ and 5.1$\sigma$ significance levels are $4.2\times 10^{-6}$ and $3.4 \times 10^{-7}$, respectively. (We assume that test statistics are distributed as $N(0, 1)$ under the null for two-sided tests.) Therefore, the largest index satisfying $p_i< (0.0027\times i) / 2$ is 2, leading to the rejection of the null hypothesis in both tests. Even though both FDR and FWER end up with the same rejection results 
in this example, 
%\dvd{in this example}, 
it is worth noting that the former is 
ensuring that
%\dvd{ensuring that} 
%controlling 
the FDR 
is
%\dvd{is} 
less than or equal to 0.0027 while the latter is controlling the FWER.

%\section{Gaussian Approximations are Generally Accurate but Sparse Observations Needs More}
\section{All model checks consider variations of the data, but some variants are more relevant than others.}\label{maxim6}

The notion of replicate data or repeated experiments is fundamental to frequentist statistical methods. In hypothesis testing, for example, evidence is quantified by (mathematically or numerically) computing the distribution of the test statistic that would result if multiple replicate data sets were generated under the null. In Bayesian data analysis, on the other hand, the posterior predictive distribution %is the distribution 
of additional data given the observed data
%predictive distribution conditioning on the entire observed data, are 
%It 
is used to generate replicate data sets. %The replicate datasets 
In both cases, the replicate data represent the statistical variability and possible range of a test/summary statistic, %due to  noise in the data, 
and are used to quantify the expected deviation between the observed data and the null/posited model, thus enabling researchers to quantify uncertainty.%, that is, \emph{what they do not know.}

At first blush generating replicates may seem to be a well-stipulated proposition. In practice, however, researchers must consider how the replicates should be generated to be most comparable with their real data. For example, a researcher may only wish to consider replicate data with the same experimental conditions, instrumental effects, exposure time, and sample size as their real data.  These are known quantities; %that we should \emph{not lose track of} by and 
varying them among the replicate data sets %, which 
can make our uncertainty quantification less relevant for the actual uncertainty we care about. Conditioning on these factors reduces the variability of the replicate data sets and makes them more comparable with the real data. This in turn reduces uncertainty, error bars on fitted parameters, and the lengths of confidence intervals; similarly it increases the statistical power to distinguish between the null and the alternative in a hypothesis test.

Such considerations have led to the broad emphasis in statistics on conditioning as much as feasible; see Section~5.2 of \cite{craiu2023six} for a succinct overview. It has also led statistical theorists to consider if there is flexibility to condition on further attributes of the data in order to further increase statistical power. In a goodness-of-fit test, for example, the aim is to quantify the deviation between the observed data and the fitted model and to assess if it is greater than would be expect under the null. It seems entirely appropriate in this setting to only consider replicate data that have the same fitted model as the real data, e.g., by conditioning on the fitted model parameters\footnote{Monte Carlo simulations in astronomy, for instance, obtain uncertainties of unknown parameters by generating replicated data sets given the maximum likelihood estimate computed on the observed data, fitting the model on each replicate set again, and quantifying  the variations of the estimated parameters \citep[e.g.,][]{tewes2013b}.}. Such a procedure is expected to reduce variability among the replicates (as they all have the same fitted parameters), make them more comparable with the real data, and increase statistical power. 
In this case, $p$-values are typically obtained via the parametric bootstrap~\citep{efron1985bootstrap}, where the estimated parameters are used as the ``ground truth parameter'' when generating replicate data sets.

\citet{wood} consider the specific example of 
%from high-energy physics that is pertinent to astrophysics, that of a 
background contaminated Poisson counts. Letting the observed count $Y^{\rm obs}$ equal the sum of the unobserved source, $Y_S$, and background $Y_B$ counts, \citet{wood} make that astute observation that while $Y_B$ is unknown, it is bounded by $Y^{\rm obs}$, i.e., we know $Y_B\leq Y^{\rm obs}$. By considering only replicates data with  $Y_B^{\rm rep} \leq Y^{\rm obs}$, \citet{wood} devise more coherent confidence intervals for the source intensity.

To give a concrete example of the advantage of conditional goodness-of-fit tests, we consider a Poisson model, where
\begin{equation}
N_i \ \indep \  {\rm Poisson} (s_i(\boldsymbol{\theta})) 
\label{eq:pois}
\end{equation}
is the count in bin $i$ for $i=1,\ldots, 10$. The test compares a uniform null model, where all of the bins have the same expected count, $\lambda$, with the fully saturated alternative model where each bin has its own expected count, $s_i$: 
%We give another example that showcases the advantage of using conditional distributions to create replicates for goodness-of-fit tests for a simplified example in X-ray spectra analysis. 
%In this example, $10$ count data, $N_i$'s ($i=1, 2, \ldots, 10$), are modeled as Poisson counts of.... We test the following null and alternative hypotheses,  
\begin{align}\nonumber
    \begin{aligned}
        H_0:&~~s_i(\boldsymbol{\theta}) = \lambda~~\\
        H_a:&~~s_i(\boldsymbol{\theta}) = s_i.
    \end{aligned}
\end{align}
We use the $C$-statistic \citep{cash1979, kaastra2017use}, defined as minus twice the logarithm of the ratio of the likelihood under the null and that under the alternative, %of the parametric model $\{N_i\sim {\rm Poisson}(s_i(\boldsymbol{\theta}))$ to that of the fully saturated model $\{N_i\sim {\rm Poisson}(s_i)$, 
with both likelihoods  evaluated at their respective maximum likelihood estimates. Specifically, the maximum likelihood estimate under the null is $\hat\lambda = \sum_{i=1}^{10} N_i/10$ and under the alternative is $\hat s_i = N_i$. 
%We consider a special case where $s_i(\boldsymbol{\theta}) = \lambda$, the same for all $i$, thus the maximum likelihood estimate for $\lambda$ is $\sum_{i=1}^{10} N_i/10$. Consequently, the C-statistic depends on the maximum likelihood estimate and the observed data.  
%We note that 
We consider two null distributions for the $C$-statistic. The \emph{unconditional null} resamples data according to the Poisson$(\hat{\lambda})$ distribution. The \emph{conditional null}, on the other hand, conditions on the maximum likelihood estimate of $\lambda$ %, multiplied by $n$, which 
which is equivalent to conditioning on the the total count, $\sum_{i=1}^n N_i$, resulting in resampling data from a multinomial distribution.

A simulation study demonstrates that the power of this goodness-of-fit test (i.e., the probability of correctly rejecting the null hypothesis) can increase by more than 30\% when the significance level is set to 0.0027 (corresponding to the typical 3$\sigma$ threshold in astronomy); see Appendix \ref{appendix} for more details. \citet{li2024cstat}  provide a rigorous study of conditional and unconditional goodness-of-fit tests based on the $C$-statistic, under a general framework designed for  realistic high-energy spectral models.
%the percentage improvement from using the conditional null distribution exceeds 
%, denoted by the dot-dashed vertical line, the percentage improvement from using the conditional null distribution exceeds 30\%

\section{Concluding Remarks}
Astronomical data are now being produced at an unprecedented rate 
and
%\dvd{and} 
with increasing complexity, and even more large-scale telescopes are expected to come into operation soon. Even though the quantity and complexity of modern astronomical data naturally demand sophisticated statistical tools for various purposes,  no single all-purpose statistical tool exists that can be deployed without careful consideration of its limitations and underlying assumptions. Rather state-of-the-art statistical methods require care, both in selecting an appropriate method and applying it properly. In some cases, existing methods do not suffice and new techniques must be developed. All together, this means that astronomers must be cognizant of the limitations and assumptions of the statistical and machine learning tools they employ, and must be cautious when using them.

We have proposed six statistical maxims to promote statistically sound data analytic practices and to improve the quality of scientific findings  in astronomy.   We hope that researchers are able to easily check these maxims as part of their daily data analytic routines. These maxims, however, are certainly not sufficient to solve all possible problem that might arise from the myriad of data types used in astronomical data analyses. 
For example,  as a reviewer pointed out,  one may question ``whether our maxims, or any other broad-sweep approaches towards reliability of statistical conclusions, apply to machine learning methods that have either only an algorithmic foundation or can be viewed as models with vast number of parameters.''

Whereas several of our maxims are generally applicable to any empirical studies, such as the first and second maxims, we echo this reviewer's call for someone to lead `Six Maxims for Applying Machine Learning to Astronomy.' More broadly, we hope our work will encourage experts in other components of the data life cycle, such as data management and data visualization \citep[see][]{Wing2019Data}, to develop their own Six Maxims to benefit the astronomy community, as it continuously improves its ability to extract scientific value from complex astronomical data.
%More broadly, we hope our work can encourage experts dealing with other components of the data life cycle, such as data management and data visualizations  to provide their Six Maxims to benefit the astronomy community as it continuously improves its ability to extract scientific value out of complex astronomical data.} 

\phantom{Whereas several of our maxims are generally applicable to any empirical studies}
\acknowledgments

This work was conducted under the auspices of the CHASC International Astrostatistics Center. CHASC is supported by NSF grants DMS-18-11308, DMS-18-11083, DMS-18-11661, 
DMS-21-13615, DMS-21-13397, and DMS-21-13605; by the UK Engineering and Physical Sciences Research Council [EP/W015080/1];
%This work was conducted under the auspices of the International CHASC Astrostatistics Center. CHASC is supported by NSF DMS-18-11308, DMS-18-11083, and DMS-18-11661, 
and by NASA APRA 80-NSSC21-K0285. We thank Eric Feigelson for initiating discussion and for providing insightful comments, which led to this work.  We thank CHASC members for many helpful discussions, the graduate students who took the Stat 303 class at Harvard University in 2023 for reading and evaluating an earlier version of the manuscript, and two anonymous referees for their thoughtful comments. HT extends deep appreciation to Jogesh Babu for sharing his invaluable insights into the statistical challenges in astronomy and for offering his helpful guidance during the preparation of this manuscript. DvD's work was supported in part by a Marie Sk{\l}odowska-Curie RISE Grants (H2020-MSCA-RISE-2015-691164, H2020-MSCA-RISE-2019-873089)  provided by the European Commission.  VLK and AS acknowledge support from NASA Contract NAS8-03060 to the Chandra X-ray Center. KSM is supported by the European Union’s Horizon 2020 research and innovation programme under ERC Grant Agreement No.~101002652 and Marie Sk{\l}odowska-Curie Grant Agreement No.~873089. YC acknowledges support from NASA 22-SWXC22\_2-0005, NASA 22-SWXC22\_2-0015, and NSF PHY 2027555.

\appendix

\begin{figure}[t!]
\begin{center}
\includegraphics[width=3in]{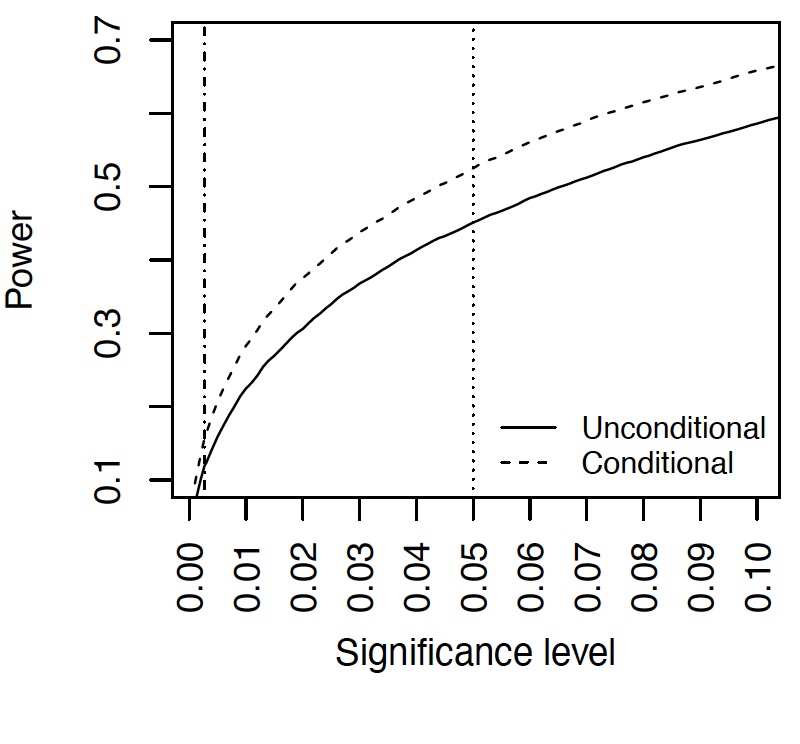}\\
\includegraphics[width=3in]{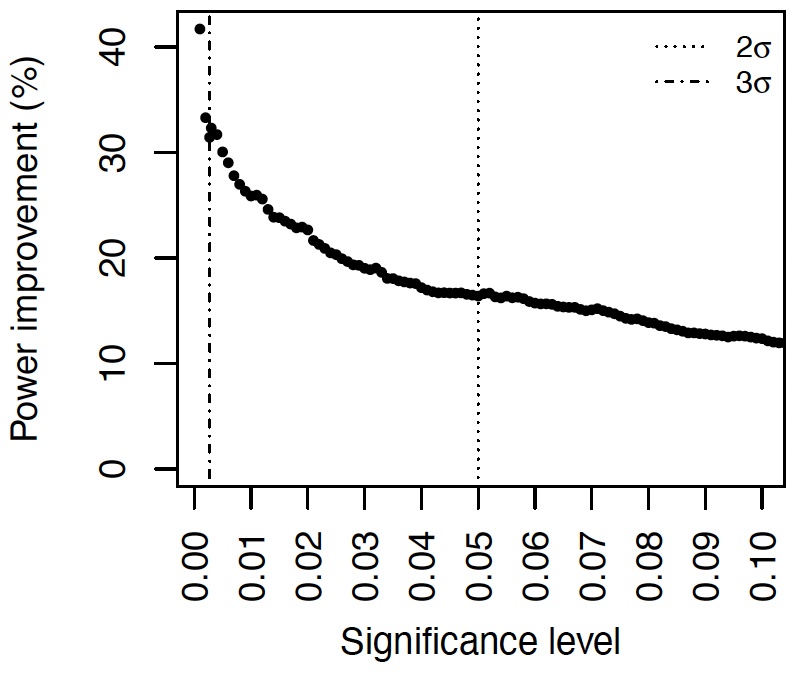}
\caption{Upper panel:  Comparison of the power (i.e., the probability of rejecting the null hypothesis when it is false) of a conditional (dashed curve) and unconditional (solid curve) goodness-of-fit test based on the $C$-statistic. % using the conditional null distribution (blue) and the unconditional distribution (red). 
We compare the uniform null model given in (\ref{eq:pois}) with $s_i(\boldsymbol{\theta}) = \lambda$ to a particular alternative model with $s_i(\boldsymbol{\theta})$ varying linearly between $2$ and $9$. The power obtained from the conditional test is uniformly higher than that of the unconditional test. The dot-dashed vertical line indicates the significance level of the $3\sigma$ threshold (i.e., 0.0027), while the dotted vertical line represents the $2\sigma$ threshold (0.05).
Bottom Panel: Percentage improvement in power as a function of the statistical significance level. With the $3\sigma$ threshold, indicated by the dot-dashed vertical line, the power improves by $31.4\%$. At the $2\sigma$ threshold, denoted by the dotted vertical line, the power improves by $16.4\%$.}
\label{fig:condi}
\end{center}
\end{figure}

\section{A simulation study: Conditional vs unconditional goodness-of-fit tests}\label{appendix}

We conduct a simulation study to illustrate the advantage of the conditional test in terms of statistical power  (i.e., the probability of correctly rejecting the null hypothesis) against a particular alternative model, namely the model in Eq.~(\ref{eq:pois}) with $s_i(\boldsymbol{\theta})$ varying linearly from 2 to 9 across the ten bins.  We independently simulate 20,000 data sets under this alternative Poisson model. For each simulation, we compute $\hat\lambda^{(j)}$ for $j=1, 2, \ldots, 20000$, and then simulate a further 5,000 data sets from each of  (i) the unconditional null distribution (Poisson with $\hat\lambda^{(j)}$) and (ii) the conditional null distribution (multinomial with a total of $10\times\hat\lambda^{(j)}$ counts). This allows us to numerically compute the $p$-value associated with the $C$-statistics from each of the 20,000 simulated data sets, and the power of the conditional and unconditional tests as the proportion of these $p$-values that are less than a given significance level (i.e., the probability of incorrectly rejecting the null).

The results are shown in Figure~\ref{fig:condi}.  The upper panel illustrates that the power obtained from the conditional test (denoted by the dashed curve) is uniformly greater than that of the unconditional test (represented by the solid curve). To emphasize this improvement, the percentage increase in power is displayed in the bottom panel. Although the percentage improvement decreases as the significance level increases, it remains at least 10\% when the significance level is below 0.1.  In particular, when the significance level is set to 0.0027 (corresponding to the typical 3$\sigma$ threshold in astronomy), denoted by the dot-dashed vertical line, the percentage improvement from using the conditional null distribution exceeds 30\%.

%, \hl{especially ANYONE??}
\bibliography{maxims}{}
\bibliographystyle{aasjournal}

\end{document}